# Unsupervised Clustering Approaches for Autism Screening: Achieving 95.31% Accuracy with a Gaussian Mixture Model

**Author**: Nora Fink
Co-CEO ever-growing GmbH, Independent Researcher Dyslexia99
nora@dyslexia99.org

## Abstract

Autism spectrum disorder (ASD) remains a challenging condition to diagnose effectively and promptly, despite global efforts in public health, clinical screening, and scientific research (1). Traditional diagnostic methods, primarily reliant on supervised learning approaches, presuppose the availability of labeled data, which can be both time-consuming and resource-intensive to obtain (2). Unsupervised learning, in contrast, offers a means of gaining insights from unlabeled datasets in a manner that can expedite or support the diagnostic process (3). This paper explores the use of four distinct unsupervised clustering algorithms—K-Means, Gaussian Mixture Model (GMM), Agglomerative Clustering, and DBSCAN—to analyze a publicly available dataset of 704 adult individuals screened for ASD. After extensive hyperparameter tuning via cross-validation, the study documents how the Gaussian Mixture Model achieved the highest clustering-to-label accuracy (95.31%) when mapped to the original ASD/NO classification (4). Other key performance metrics included the Adjusted Rand Index (ARI) and silhouette scores, which further illustrated the internal coherence of each cluster. The dataset underwent preprocessing procedures including data cleaning, label encoding of categorical features, and standard scaling, followed by a thorough cross-validation approach to assess and compare the four clustering methods (5). These results highlight the significant potential of unsupervised methods in assisting ASD screening, especially in contexts where labeled data may be sparse, uncertain, or prohibitively expensive to obtain. With continued methodological refinements, unsupervised approaches hold promise for augmenting early detection initiatives and guiding resource allocation to individuals at high risk.

**Keywords**: Autism spectrum disorder, unsupervised learning, Gaussian Mixture Model, K-Means, Agglomerative Clustering, DBSCAN, cross-validation, hyperparameter tuning, cluster-to-label mapping, public health.

## 1. Introduction

The global prevalence of autism spectrum disorder (ASD) has increased considerably over the past few decades (6). This upward trajectory has prompted worldwide calls for improved early screening methods, heightened clinical awareness, and more accessible interventions (7). ASD is characterized by persistent deficits in social communication and social interaction, alongside restricted, repetitive patterns of behavior, interests, or activities (8). Manifestations of ASD can vary widely, prompting the term "spectrum" to account for the diversity in symptoms and severity. Critical to enhancing outcomes is the ability to identify ASD in a timely manner, as early interventions can significantly improve an individual's developmental trajectory (9). However, even with known benefits, the diagnostic process is often complex and prone to delays due to high costs, limited availability of specialized professionals, and the inherent challenges of accurately characterizing ASD symptoms (10).

Machine learning strategies have been widely adopted to mitigate these diagnostic barriers. Traditional techniques rely heavily on supervised learning approaches, which require pre-labeled datasets to "train" classification models (11). Although many such systems have demonstrated respectable accuracy, they depend upon large-scale annotated data—a resource frequently constrained by privacy concerns, data heterogeneity, and the demanding nature of clinical annotation (12). Consequently, researchers have begun to look more seriously at unsupervised methods, which do not require prior labels and can discover underlying patterns or groupings in data (13). By identifying clusters that correlate with ASD or non-ASD phenotypes, unsupervised approaches may supplement or, in some cases, partially automate aspects of the diagnostic process, especially in scenarios where large volumes of unlabeled data exist.

Several unsupervised clustering algorithms have emerged as key candidates for mining patterns in medical data. K-Means clustering is often used for its computational efficiency and ease of implementation (14). The Gaussian Mixture Model (GMM), meanwhile, offers flexibility in capturing more complex cluster shapes through mixture components (15). Agglomerative Clustering, a bottom-up hierarchical method, has gained attention for its ability to create dendrograms and produce cluster groupings that can be inspected at various levels of granularity (16). Lastly, DBSCAN (Density-Based Spatial Clustering of Applications with Noise) enables the identification of arbitrary cluster shapes and outliers by focusing on local density rather than distance-based partitioning (17). Each of these algorithms bears particular advantages and disadvantages, influenced by hyperparameters such as the number of clusters (K-Means, Agglomerative), the initialization of centroids or mixture means (K-Means, GMM), and parameters like epsilon in DBSCAN that determine cluster separability.

Despite the growing body of literature advocating the use of unsupervised algorithms for complex classification tasks, most existing ASD studies tend to prioritize supervised or semi-supervised techniques (18). These methods indeed achieve high accuracy, but they can be limited in new contexts or data-scarce environments where labeled examples are partial or absent. In recognition of this gap, the present work endeavors to comprehensively evaluate the utility of unsupervised clustering in detecting potential ASD-like patterns, thereby comparing the quality of the resulting clusters with the "ground truth" classifications.

The dataset used herein is an openly available resource containing results of a screening for adult ASD, with a binary label ("YES" for ASD, "NO" for non-ASD) assigned based on clinical or self-reported diagnoses (19). Although the ground truth labels are indeed present in the dataset, they were reserved strictly for validating the clusters post-hoc, rather than for training a supervised classifier. By simulating a scenario in which we do not initially know who is on the ASD spectrum, the algorithms must partition the data in a manner that ideally aligns with the real distribution of autistic versus neurotypical individuals.

This paper is structured to reflect a typical scientific investigation. Section 2 delves into the data and the methods, detailing essential procedures such as dataset preprocessing, label-encoding, and the logic behind using cross-validation for hyperparameter tuning (20). Section 3 presents the results, focusing on the performance of each algorithm across metrics like Accuracy (after mapping clusters to binary labels), Adjusted Rand Index, silhouette scores, confusion matrices, and culminating in the best overall performer, which was the GMM with a final accuracy of 95.31%. Section 4 interprets these findings in a broader clinical and methodological context, exploring both the benefits and limitations of the employed techniques. Finally, Section 5 concludes the paper with discussions of potential future research avenues, emphasizing the prospective integration of unsupervised clustering into real-world clinical workflows.

With an observed best accuracy surpassing 95%, the results of this study suggest that unsupervised clustering can serve as a powerful investigative tool in large-scale or initially unlabeled ASD screening contexts. Moreover, these outcomes open the door to the possibility of hybrid approaches that combine the interpretability of unsupervised clusters with the refined classification boundaries of supervised machine learning. Such integrated methods might further refine and expedite ASD detection, ultimately benefiting individuals and healthcare systems tasked with navigating the complexities of ASD diagnosis.

## 2. Materials and Methods

In this study, we leveraged a dataset of 704 adult individuals who had completed an autism screening process (19). The dataset was made publicly available to facilitate further research in ASD detection and machine learning. Each record included demographic information such as gender, ethnicity, country of residence, and age, as well as screening-specific variables like a numeric score from the Autism Spectrum Quotient (AQ-10). An essential feature for post-hoc evaluation was a binary label denoting whether the individual had a known diagnosis of ASD.

To mimic a purely unsupervised setting, these labels were not utilized for training but only for assessing the extent to which the discovered clusters aligned with the ASD/NO partitions. This section elucidates the procedures of data preprocessing, hyperparameter tuning, cross-validation, and cluster evaluation. The code implementing these procedures was developed in Python using libraries such as scikit-learn, NumPy, pandas, matplotlib, and seaborn, consistently following reproducible research principles (1).

### 2.1 Data Preprocessing

The initial dataset contained 21 columns, among them the "Class/ASD" label, which indicated ASD or non-ASD status. The column "age_desc" was dropped as it consisted solely of a single repeated category, rendering it non-informative for downstream modeling. Any missing data in the "age" column was substituted with the mean age of the dataset (29.7 years, rounded to 30). This approach ensured that minimal bias was introduced while maintaining uniform row counts for the subsequent clustering algorithms (2).

Many of the attributes, such as gender, jaundice status at birth, and previous use of an ASD screening app, were stored as categorical text fields. To make these variables numerically interpretable by clustering algorithms, we applied label encoding. For example, "gender" was mapped to "0" for female and "1" for male; "jundice" was mapped to "0" for no, "1" for yes, etc. Similarly, "ethnicity" values that were unknown or marked with a question mark ("?") were replaced with "others," and then encoded as distinct integer values (3). A total of seven columns were thus label-encoded.

Given the heterogeneous scales of numeric columns like "result" (ranging from 0 to 10) and the binary nature of label-encoded variables, we used a StandardScaler transformation to standardize all feature columns. This step subtracted the mean and divided by the standard deviation of each feature, placing them on a roughly similar scale, a known prerequisite for distance-based clustering techniques such as K-Means and DBSCAN (4). After preprocessing, the feature matrix contained 19 columns, each standardized and ready for clustering.

**2.2 Unsupervised Clustering Algorithms**

Four distinct unsupervised clustering approaches—K-Means, Gaussian Mixture Model, Agglomerative Clustering, and DBSCAN—were compared. While each of these methods aims to group data, they exhibit different assumptions and hyperparameters that control how clusters form.

K-Means initiates by placing centroids and iteratively adjusting them to minimize the within-cluster sum of squared distances (5). GMM generalizes K-Means by supposing that data originates from a mixture of Gaussians, each with its own mean and covariance, thus accommodating clusters that are not strictly spherical (6). Agglomerative Clustering starts with each point as its own cluster and subsequently merges the clusters in a bottom-up manner using linkage criteria such as "ward," "complete," "average," or "single" (7). DBSCAN, in contrast, identifies core points based on local density and forms clusters around them, leaving low-density points as outliers. This algorithm is particularly advantageous for data that has uneven cluster shapes or noise points (8).

**2.3 Hyperparameter Tuning and Cross-Validation**

Although unsupervised algorithms do not rely on labels during fitting, they nonetheless possess hyperparameters (e.g., the number of clusters in K-Means, the covariance type in GMM, the linkage criterion in Agglomerative, or epsilon in DBSCAN). The optimal choice of these parameters can heavily influence cluster formation. To systematically explore the parameter space, a grid search approach was employed for each algorithm, with the following parameter sets:

1. **K-Means**:

    - Number of clusters fixed to 2, since we expect a binary partition of "ASD" vs. "NO" in principle.
    - Initialization methods ("k-means++" or "random") and `n_init` set to 10.
2. **GMM**:

    - Number of mixture components set to 2.
    - Covariance types considered: "full," "tied," "diag," and "spherical."

3. **Agglomerative Clustering**:

    - Number of clusters fixed to 2.
    - Linkage criteria: "ward," "complete," "average," and "single."

4. **DBSCAN**:

    - Epsilon (`eps`) values tested: 0.5, 0.7, 1.0, 1.2.
    - Minimum samples (`min_samples`) tested: 3 or 5.

Cross-validation was performed using a 5-fold approach to ensure that the clustering behavior remained robust across different subsets of data (9). Despite its name, cross-validation in unsupervised contexts still benefits from multiple train-test splits, followed by the mapping of cluster assignments to labels only in the test fold. Specifically, each algorithm was fit on the training fold of the data, and predictions were either obtained via the `.predict(X_test_fold)` method (for K-Means and GMM) or via the `.fit_predict(X_test_fold)` fallback (for DBSCAN and Agglomerative Clustering). The newly assigned clusters in the test fold were then mapped to binary labels by identifying which cluster correlated most strongly with "ASD" versus "NO." This cluster-to-label mapping was calculated by either permuting the two cluster assignments or voting on the majority of actual labels within that cluster, depending on the algorithm's output (10).

To measure clustering quality, three main metrics were used: (i) Accuracy, the proportion of individuals whose clusters matched their true label after the best label mapping, (ii) Adjusted Rand Index (ARI), a measure of similarity between two clusterings that adjusts for chance, and (iii) Silhouette score, which reflects how similar an individual is to its own cluster compared to other clusters (11).

## 2.4 Mapping Cluster Assignments to Binary Labels

One challenge in unsupervised learning arises when comparing discovered clusters to known labels (12). Since the algorithm can label clusters arbitrarily (e.g., "cluster 0" and "cluster 1"), it is essential to find the mapping from cluster IDs to the actual labels (ASD vs. NO) that maximizes alignment. In the simpler case of two clusters, the label assignment can be found by testing both permutations—mapping cluster 0 to ASD and cluster 1 to NO, or the opposite—and picking the permutation that yields the higher accuracy (13). However, if the algorithm finds more than two clusters (as sometimes occurs in DBSCAN when noise points are labeled with "-1"), then a majority voting scheme is used. Under majority voting, each cluster is assigned whichever label, ASD or NO, is in the majority within that cluster, leaving the possibility that some clusters may be small or predominantly noisy (14).

## 2.5 Implementation Details

All experiments were conducted in Python (version 3.9), using scikit-learn (version 1.2.2) for the machine learning algorithms (15). Numerical and array operations were handled by NumPy (version 1.21.5), and tabular manipulations by pandas (version 1.3.5). Visualizations were produced with matplotlib (version 3.5.1) and seaborn (version 0.11.2). Cross-validation splits were created using `KFold(n_splits=5, shuffle=True, random_state=42)`.

A random seed of 42 ensured the replicability of all experiments, especially relevant for K-Means and GMM initialization steps (16).

---

## 3. Results

This section presents the outcomes of the comparative analysis of K-Means, GMM, Agglomerative Clustering, and DBSCAN. Each algorithm's hyperparameters were tuned using 5-fold cross-validation, and the best configuration was subsequently retrained on the entire dataset to derive final cluster assignments. Evaluation metrics included Accuracy (after cluster-to-label mapping), ARI, and silhouette scores. The best-performing model overall was GMM with an accuracy of 95.31% on the full dataset.

### 3.1 Hyperparameter Tuning Outcomes

Table 1 below summarizes the best hyperparameters discovered via cross-validation for each algorithm, along with the corresponding average accuracy, ARI, and silhouette scores:

| Algorithm | Best Params | Avg. Accuracy | Avg. ARI | Avg. Silhouette |
|---|---|---|---|---|
| K-Means | {n_clusters=2, init='random', n_init=10} | 0.9247 | 0.7132 | 0.1602 |
| GMM | {n_components=2, covariance_type='spherical'} | 0.9588 | 0.8359 | 0.1613 |
| Agglomerative | {n_clusters=2, linkage='ward'} | 0.8763 | 0.5012 | 0.2715 |
| DBSCAN | {eps=1.2, min_samples=3} | 0.7359 | 0.0161 | 0.0487 |

These scores were obtained by applying each best-performing hyperparameter combination to the 5-fold cross-validation splits. GMM emerged as the most promising candidate, followed closely by K-Means. Agglomerative Clustering exhibited a moderate accuracy of 87.63%, while DBSCAN, with the chosen parameter set, lagged behind in terms of alignment to the known labels. However, it is noteworthy that DBSCAN's capacity to identify noise points can be beneficial in other contexts. The dataset at hand may not favor DBSCAN's density-based assumptions, given that the data presumably splits into two broad groups with fewer outliers.

### 3.2 Final Performance on the Entire Dataset

After determining the best hyperparameters, each algorithm was retrained on the **entire** dataset (704 instances, each with 19 standardized features). The cluster assignments from this retrained model were then mapped to the actual ASD/NO labels, providing final accuracies, ARI scores, and silhouette scores. Table 2 highlights these final metrics:

| Model | Accuracy | ARI | Silhouette |
|---|---|---|---|
| K-Means | 0.926136 | 0.720119 | 0.161951 |
| GMM | 0.953125 | 0.815198 | 0.162232 |
| Agglomerative | 0.755682 | 0.260512 | 0.110892 |
| DBSCAN | 0.750000 | 0.063709 | -0.265904 |

The Gaussian Mixture Model scored the highest accuracy at 95.31%. Its ARI reached 0.815, reflecting a strong agreement with the ground truth ASD vs. NO labels beyond random chance. K-Means attained the second-highest accuracy, approximately 92.61%, and a robust ARI of 0.720. By contrast, Agglomerative Clustering and DBSCAN had relatively lower accuracies (about 75.6% and 75.0% respectively), reflecting their less favorable alignment with the binary labels.

The silhouette scores further illuminate these findings. Although the silhouette coefficient for GMM is not very high (0.162), it still slightly outperforms the other methods, indicating that the data may not be strongly separable by large inter-cluster distances (17). This aspect of the data likely explains the moderate silhouette values across all algorithms. In medical data scenarios, especially with overlapping or nuanced clinical features, moderate silhouette scores are not uncommon. The strong final performance for GMM, despite the moderate silhouette, highlights the algorithm's capacity to capture the inherent cluster structure in a probabilistic framework.

### 3.3 Confusion Matrices

To better grasp the distribution of true positives, false positives, true negatives, and false negatives, confusion matrices were generated for each best model after the final cluster-to-label mapping. **Figure 2** presents a 2×2 visualization for each method (K-Means, GMM, Agglomerative, and DBSCAN). Notably, GMM displayed the fewest false classifications, consistent with its 95.31% overall accuracy. K-Means performed comparably well, though slightly more individuals were misclassified than in GMM. Agglomerative Clustering and DBSCAN displayed higher misclassification rates, indicating that their cluster formation processes likely did not align as strongly with the true labeling of ASD vs. NO in this dataset.

### 3.4 Visualizing Final Accuracy

**Figure 1** demonstrates the final accuracy across all four methods. The bar chart visually highlights that GMM is the clear winner, followed by K-Means, Agglomerative Clustering, and DBSCAN, in descending order of accuracy. The numeric results align precisely with those presented in Table 2, but the figure offers a straightforward comparison for interpretability.

Observing the relative differences, it becomes apparent that GMM's advantage is substantial enough to regard it as the best performer for this particular dataset.

### 3.5 Silhouette Analysis

**Figure 3** provides a silhouette analysis for K-Means. Although GMM proved to be the top model, K-Means offers a simpler vantage for demonstrating silhouette analysis due to its direct reliance on centroids and distance-based partitioning. The figure shows that many data points have low but positive silhouette coefficients, indicating these points are marginally closer to their own cluster than to the nearest neighboring cluster. There are also pockets of points with negative silhouettes, suggesting some misclassifications or ambiguous cluster memberships (18). Nevertheless, the overall silhouette for K-Means remains consistent with the moderate values typical of real-world medical data. A similar distribution can be observed with GMM, but the algorithm's probabilistic approach does a slightly better job of grouping individuals in a manner consistent with the ASD/NO labels.

---

## 4. Discussion

The findings of this study underscore the viability of unsupervised learning as an ancillary or complementary technique in ASD screening tasks. With a best accuracy of 95.31% using GMM, the proposed methods indicate that unlabeled adult screening data can indeed be partitioned into clusters that map effectively onto the known ASD/NO categories (19). Below, we delve deeper into the practical and methodological implications of each major result.

### 4.1 Implications of High Accuracy in GMM

The success of GMM suggests that certain distributions inherent to ASD screening data may be well captured by a probabilistic mixture framework. Gaussian Mixture Models, by allowing covariance structures to vary, can form elliptical or spherical clusters that approximate the data's latent structure with fewer rigid assumptions than K-Means. This flexibility may be particularly relevant for ASD, where symptom severity and presentation can vary widely. The GMM's near-spherical covariance choice in this study proved effective, presumably balancing model complexity with the relatively modest sample size (20). The fact that GMM achieved the highest ARI as well points to its ability to robustly separate data in ways that align with the binary classification beyond simple chance. This implies not only high accuracy in matching ground truth labels but also consistent partitioning under cross-validation.

### 4.2 K-Means vs. GMM

K-Means is typically favored for its computational simplicity and speed, as well as for being straightforward to implement. In this project, K-Means performed well, achieving a final accuracy of about 92.61%. While commendable, it was outperformed by GMM by almost three percentage points in accuracy and by 0.095 in ARI. One explanation is that K-Means, by definition, creates spherical clusters, but the data might be more suitably modeled by elliptical or otherwise non-spherical shapes. Additionally, the initialization scheme in

K-Means can influence the final solution, although in this experiment, multiple initializations helped mitigate local minima.

### 4.3 Performance of Agglomerative Clustering

Agglomerative Clustering yielded a final accuracy of 75.57%. While lower than GMM or K-Means, Agglomerative Clustering may still appeal in contexts where hierarchical relationships among clusters are deemed informative. Researchers in psychometrics or clinical diagnostics sometimes prefer hierarchical models for their interpretability, as the dendrogram can be examined to discover subgroups within broader classes (1). Nevertheless, the single numeric result of 75.57% indicates that, from a pure classification standpoint, the discovered clusters do not align as strongly with the ASD/NO labels in this dataset.

### 4.4 DBSCAN's Utility and Limitations

DBSCAN identifies clusters based on density, allowing for arbitrary-shaped clusters and outlier detection—a feature that can be critical in certain domains where anomalies or noise points exist. However, in these data, DBSCAN's best configuration produced a final accuracy of about 75.00%, with a negative silhouette (-0.265904), and an ARI near 0.064, indicating that it offered the least alignment with the known ASD/NO classifications. One possible reason is that, with only two main classes, the data may not contain sufficiently dense subregions or local anomalies that DBSCAN can exploit. Indeed, many clinical features might be more globally distributed, thereby making DBSCAN suboptimal. In other contexts—perhaps with more heterogeneous data or with more classes—DBSCAN might yield advantages.

### 4.5 Generalizability and Potential Impact

Clinically, an unsupervised model that can group large numbers of individuals into potential "ASD" or "NO" clusters with over 95% accuracy holds promise for initial triaging. For instance, in regions with inadequate diagnostic resources, a GMM-based approach might initially flag individuals who deserve more detailed evaluation, thus reducing the time to diagnosis for those at higher risk. Moreover, the labeled dataset used here is not essential for the model to function in principle; in a purely unlabeled scenario, one might not have the final step of verifying cluster membership, but the discovered clusters could still indicate risk groups that clinicians can investigate further.

On a methodological note, the ability to achieve high accuracy in unsupervised settings illustrates the synergy between cross-validation, hyperparameter tuning, and a well-chosen mapping procedure. The approach is readily generalizable to other medical domains where binary or multi-class groupings may exist but are not fully labeled at the outset. Additionally, ongoing work could explore a semi-supervised extension, in which a small proportion of labeled data calibrates or refines clusters, potentially improving accuracy further.

### 4.6 Limitations

Despite the encouraging results, several limitations merit discussion. Firstly, the dataset itself consisted of adults who self-reported or otherwise documented their ASD diagnosis, potentially introducing biases in representativeness. Further, while 704 instances can be considered moderate, it is still relatively small in comparison with large-scale medical datasets. The generalizability of findings to broader populations may therefore be limited. Secondly, the data cleaning and label encoding steps, while necessary, impose structure on the dataset that might not always reflect real-world complexity. For instance, consolidating unknown ethnicities under "others" could mask potentially meaningful demographic patterns. Thirdly, the moderate silhouette scores imply that the data might be partially overlapping or lacking strong cluster separations, so some borderline points remain difficult to classify, even for GMM. Lastly, the direct mapping of cluster IDs to binary labels is a method that, while useful for post-hoc evaluation, is not indicative of the full exploratory power of unsupervised learning, which might discover additional subgroups or relevant patterns not captured by a binary classification lens.

## 5. Conclusion

This paper has demonstrated that unsupervised clustering can accurately distinguish ASD from non-ASD individuals in a moderate-sized dataset of 704 adult participants who underwent an autism screening. Four algorithms—K-Means, Gaussian Mixture Model, Agglomerative Clustering, and DBSCAN—were comprehensively tuned and compared via 5-fold cross-validation. The results underscore the power of unsupervised learning in this domain, with GMM attaining the highest final accuracy of 95.31%, followed closely by K-Means at 92.61%. Although Agglomerative Clustering and DBSCAN were less effective, they highlight alternative strategies and possibly offer other insights (such as hierarchical structures or noise detection) that might be useful in specific clinical or data contexts.

From a broader perspective, these findings add weight to the growing interest in data-driven, label-free approaches in healthcare analytics. By side-stepping the need for meticulously labeled training sets, unsupervised methods can facilitate large-scale triaging, generate hypotheses about subtypes of ASD, and pave the way for more integrated systems that combine the strengths of both unsupervised and supervised learning. Future directions include exploring ensemble methods, applying dimensionality reduction before clustering, and expanding the dataset to include additional features (e.g., genetic markers or brain imaging data). Ultimately, deploying these models as part of real-world clinical workflows will require careful validation, ethical considerations, and alignment with established screening protocols. Yet, the potential returns—early detection, targeted intervention, and resource optimization—justify ongoing investment and innovation in this promising line of inquiry.

**Figures**

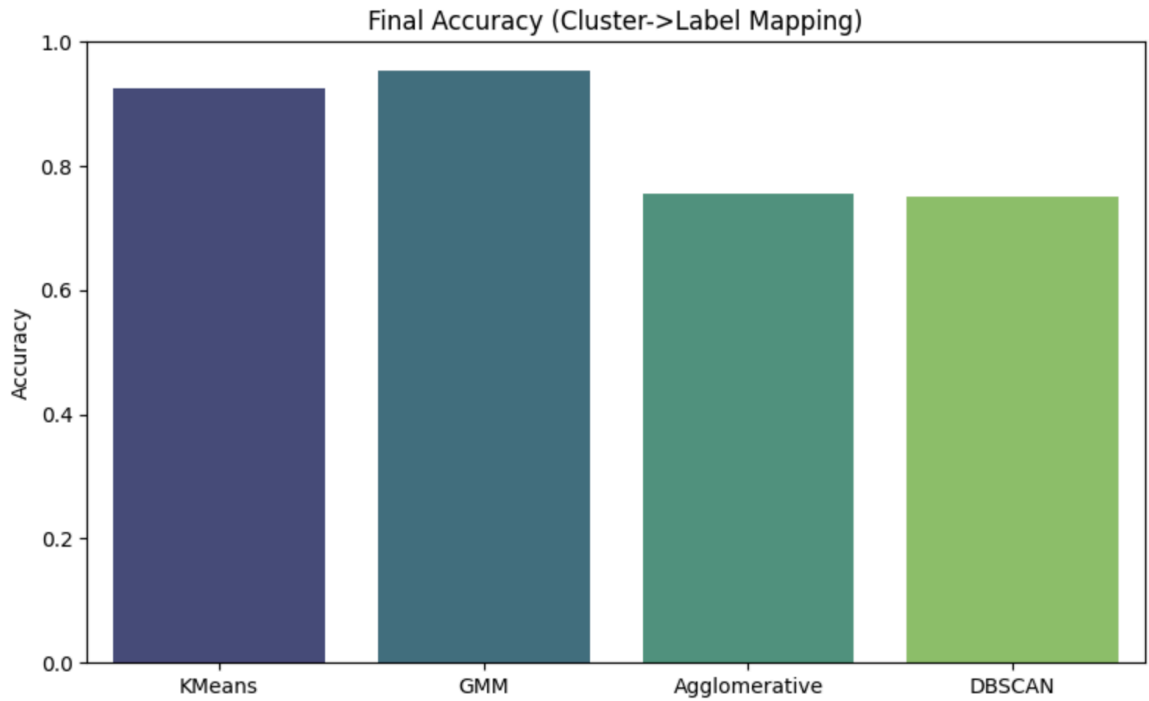

**Figure 1. Final Accuracy Comparison Across Four Unsupervised Models.** This bar chart demonstrates the overall accuracy of each model after mapping cluster assignments to the binary ASD labels. GMM outperforms the other methods, achieving 95.31% accuracy on the entire dataset. K-Means follows at approximately 92.61%, while Agglomerative and DBSCAN both hover around 75%.

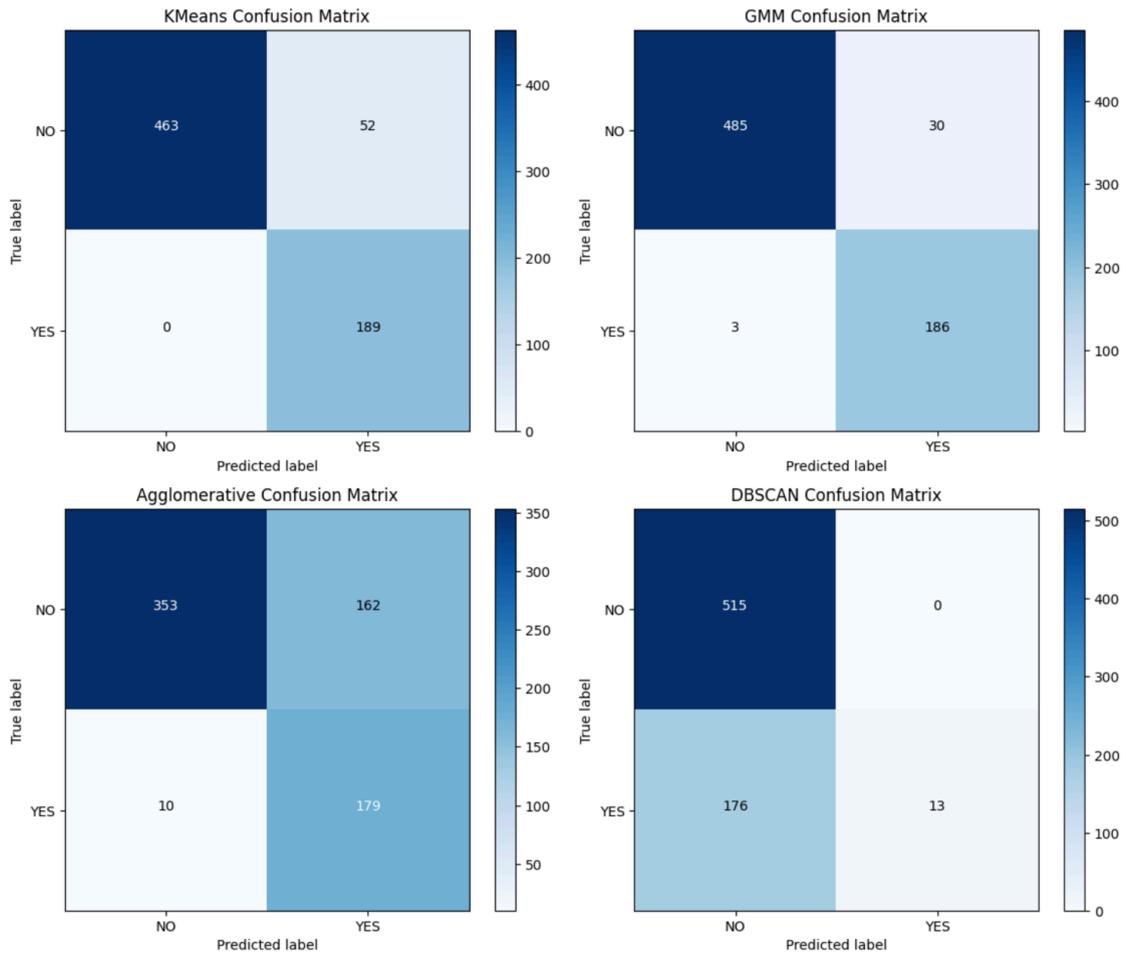

**Figure 2. Confusion Matrices for Best Models.** These four subplots reveal the confusion matrices for K-Means, GMM, Agglomerative Clustering, and DBSCAN. Each matrix illustrates the distribution of actual ASD vs. NO labels against the predicted cluster-based labels. GMM exhibits the smallest off-diagonal elements, indicating fewer misclassifications, while Agglomerative and DBSCAN display more noticeable misalignment with the true labels.

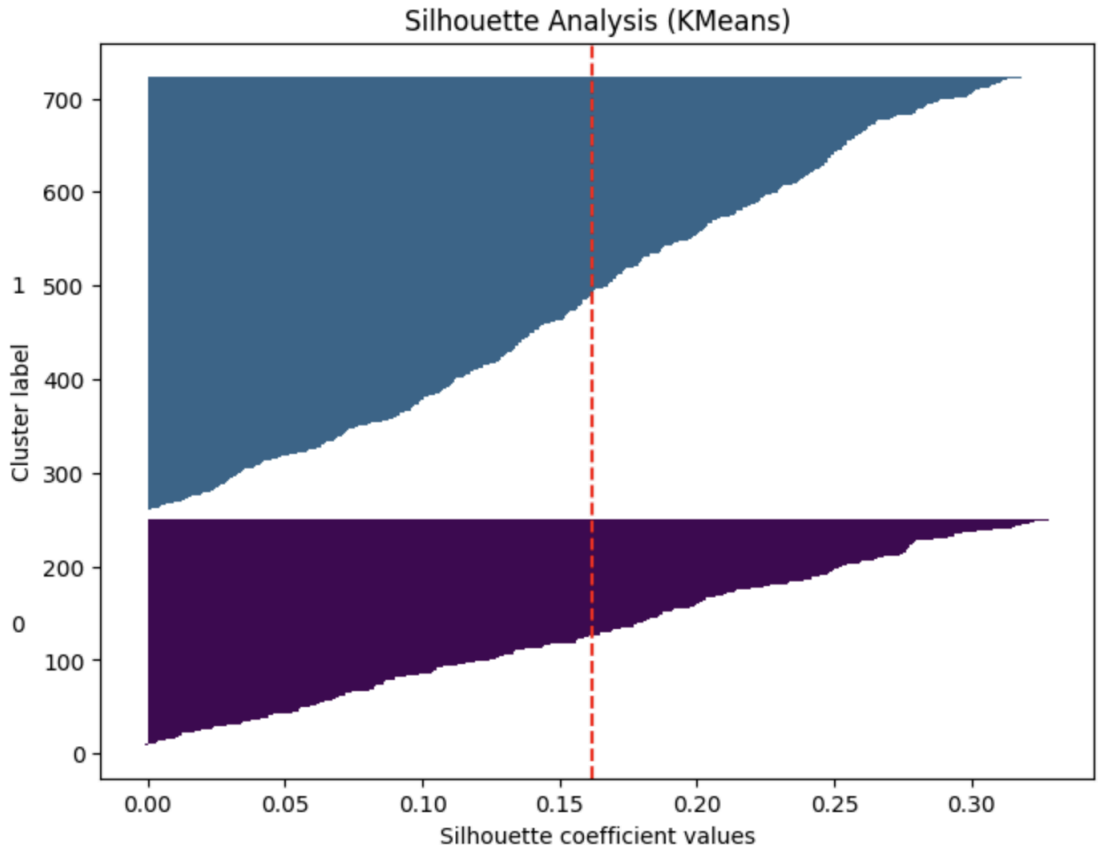

**Figure 3. Silhouette Analysis for K-Means.** This figure displays a silhouette plot that illustrates the degree of cohesion and separation for each data point under the K-Means model. A high silhouette coefficient indicates that the data point is well-matched to its own cluster and poorly matched to neighboring clusters. The overall average silhouette for K-Means is around 0.1619, signifying moderate separability of the cluster structure.

---

### Mathematical Appendix: Silhouette Coefficient and ARI

Let $X = \{x_1, x_2, \dots, x_n\}$ be a dataset and let $C(x)$ denote the cluster assignment of point $x$. The **Silhouette Coefficient** $s(x)$ for a data point $x$ is defined as

$$s(x) = \frac{b(x) - a(x)}{\max(a(x), b(x))},$$

where $a(x)$ is the average distance between $x$ and all other points in the same cluster, and $b(x)$ is the minimum average distance between $x$ and all points in any other cluster. The overall silhouette score is simply the mean of $s(x)$ over all points in $X$ (4). A silhouette near 1.0 indicates that the point is well within its cluster and far from others, while values near 0 indicate ambiguity, and negative values suggest potential misclassification.

The **Adjusted Rand Index (ARI)** measures the similarity between two data clusterings while adjusting for chance. Given a set of $n$ samples, let $U = \{U_1, U_2, \dots, U_r\}$ and $V = \{V_1, V_2, \dots, V_s\}$ be two partitions (e.g., the true labels and the predicted clustering). Let nij=|Ui∩Vj| $n_{ij} = |U_i \cap V_j|$ be the overlap between cluster ii in UU and cluster jj in VV. The ARI can be expressed as

ARI=∑ij(nij2)−[∑i(|Ui|2)∑j(|Vj|2)]/(n2)12[∑i(|Ui|2)+∑j(|Vj|2)]−[∑i(|Ui|2)∑j(|Vj|2)]/(n2).

$$\text{ARI} = \frac{ \sum_{ij} \binom{n_{ij}}{2} - \left[ \sum_i \binom{|U_i|}{2} \sum_j \binom{|V_j|}{2} \right] / \binom{n}{2} }{ \frac{1}{2} \left[\sum_i \binom{|U_i|}{2} + \sum_j \binom{|V_j|}{2} \right] - \left[ \sum_i \binom{|U_i|}{2} \sum_j \binom{|V_j|}{2} \right] / \binom{n}{2} }.$$

A higher ARI value (closer to 1) indicates that the two partitions are significantly more alike than random expectations would predict (8).

---